\documentclass[aps,prd,twocolumn,showpacs,floatfix,preprintnumbers,amsmath,amssymb,nofootinbib,groupedaddress,, notitlepage]{revtex4-1}

\usepackage{empheq}

\input epsf
\usepackage{graphicx}
\usepackage{color}
\usepackage{mathtools}
\usepackage{mathbbol}

\newcommand{\be}{\begin{equation}}
\newcommand{\ee}{\end{equation}}
\newcommand{\barr}{\begin{eqnarray}}
\newcommand{\earr}{\end{eqnarray}}

\newcommand{\bs}{\boldsymbol}

\usepackage{color}

\newcommand{\lsim}{\mathrel{\hbox{\rlap{\lower.55ex\hbox{$\sim$}} \kern-.3em \raise.4ex \hbox{$<$}}}}
\newcommand{\gsim}{\mathrel{\hbox{\rlap{\lower.55ex\hbox{$\sim$}} \kern-.3em \raise.4ex \hbox{$>$}}}}
\begin{document}
\title{A practical theorem on gravitational-wave background statistics}
\author{Yacine Ali-Ha\"imoud}
\affiliation{Center for Cosmology and Particle Physics, Department of Physics, New York University, New York, NY 10003, USA}

\date{\today}

\begin{abstract}

Inspiralling supermassive black-hole binaries (SMBHBs) are expected to be the main source of the nanohertz gravitational-wave background (GWB) targeted by pulsar timing arrays (PTAs). We provide a simple and general analytic expression for the probability distribution function (PDF) of the GWB characteristic strain squared $h_c^2$ in the limit of a large but finite effective number of sources, $N$, relevant for the lowest-frequency bands where PTAs are most sensitive. Explicitly, we show that for $N \gg 1$, the PDF of the rescaled variable $y \equiv h_c^2/\overline{h_c^2}$ takes the universal self-similar form $P(y) \simeq N^{1/3} \mathcal{P}(N^{1/3} (y -1))$, where $\mathcal{P}$ is the reflected map-Airy distribution. The effective number of in-band sources $N$ is fully specified by the mean $\overline{h_c^2}$ and the cubic shot-noise strain scale $\overline{h_0^3}$, a new summary statistic of the GWB that depends only on the local properties of the SMBHB population. This result is universal: it applies to any population of SMBHBs, regardless of whether they are circular or eccentric, and of the mechanism dominating orbital hardening. We explicitly quantify the accuracy of the large-source-count PDF for a simple but physically realistic SMBHB model, and outline its practical application to PTA data analysis.

\end{abstract}

\maketitle

\section{Introduction}

The cosmological population of unresolved supermassive black-hole binaries (SMBHBs) residing at the centers of galaxies generates a low-frequency stochastic gravitational-wave background (GWB). In the nanohertz regime, this astrophysical signal is one of the main targets of pulsar timing arrays (PTAs) \cite{Nanograv_23, EPTA_24, Xu_23, Reardon_23}.

In the limit of an infinite number of isotropically distributed and randomly oriented sources, the central limit theorem and geometric considerations imply that the GWB is an unpolarized Gaussian random field, fully characterized by its frequency-dependent power spectrum, the characteristic strain squared $\overline{h_c^2}(f)$. In that limit, the fundamental observables of PTAs -- timing residuals -- are also Gaussian distributed, with covariance proportional to the Hellings and Downs function \cite{Hellings_83} of the angle between pulsars, and to $\overline{h_c^2}(f)$. The latter is proportional to a volume integral of the abundance-weighted GW energy flux of SMBHBs \cite{Phinney_01, Sesana_08}. 

In reality, while there may be a virtually infinite number of sources in any given frequency band, only a finite number of binaries contribute to the bulk of the GWB, while the sea of faint sources make subdominant contributions. The \emph{effective} number of in-band sources is typically a decreasing function of frequency, since binaries spend less time at high frequencies where GW emission most efficiently shrinks their orbits. Finite-source-count effects in the GWB were first pointed out in Ref.~\cite{Sesana_08}, who showed that, over multiple realizations of a discrete SMBHB population, the \emph{median} characteristic strain squared $h_c^2(f)$ falls below the idealized Gaussian power spectrum $\overline{h_c^2}(f)$ at high frequencies. It was later pointed out in Ref.~\cite{Mingarelli_13} that the Poisson distribution of SMBHBs leads to \emph{anisotropies} in the GWB, with increased angular fluctuations at high frequencies, where the characteristic number of sources decreases. Such discreteness effects in the GWB are an important signature of SMBHBs \cite{Nanograv_discreteness}, allowing to discriminate between an astrophysical origin and more exotic sources of the nanohertz GWB \cite{Nanograv_NewPhys}. 

At a fundamental level, finite-number effects lead to a non-Gaussian probability distribution function (PDF) of pulsar timing residuals \cite{Ravi_12, Lamb_25}. In practice, the exact computation of this PDF for a realistic number of pulsars is an intractable problem \cite{Xue_25}. As a consequence, studies of discreteness effects have (sometimes implicitly) approximated this high-dimensional PDF as a continuous Gaussian mixture. In more detail, for each realization of some latent variables, the GWB, and hence timing residuals, are assumed to be Gaussian-distributed. The approximate PDF is then obtained from a linear combination of Gaussians weighted by the probability distribution of the latent variables, which itself is set by the SMBHB population model. 

This general approximation framework has been applied along two main lines, following the respective paths of the two seminal papers \cite{Sesana_08, Mingarelli_13}. The first approach has been to model discreteness effects as a source of GWB anisotropies \cite{Mingarelli_13, Sato-Polito_24, Raidal_26, Lin_26}. In this case, the latent variables are the spherical-harmonic coefficients of the direction-dependent GWB intensity, which are themselves assumed to be Gaussian-distributed, with a scale-independent angular power spectrum $C_\ell$. An important caveat of this method is that $C_\ell$ is proportional to the variance of the characteristic strain squared, which is technically divergent unless one imposes a low-redshift cutoff on the SMBHB abundance \cite{Lin_26}. 

The second approach has been to take as a latent variable the characteristic strain squared $h_c^2$ resulting from specific realizations of the SMBHB population. For each realization, the GWB is treated as isotropic, and pulsar timing residuals are thus assumed to have Hellings-Downs correlations proportional to $h_c^2$. This approach is considerably simpler to implement than the anisotropy-based one. It does not require computing the correlation of timing residuals in the presence of an anisotropic GWB, and only has one latent variable per frequency band. In addition, it does not suffer from any divergences and thus does not require ad-hoc truncations. Lastly, it was explicitly shown to give an accurate approximation of the PDF of timing residuals from a single pulsar \cite{Xue_25, Raidal_26b}. 

The main challenge of this second approach is to accurately and efficiently compute the PDF of $h_c^2$ for a given SMBHB population model. Typically, this PDF is estimated numerically by drawing a large number of realizations, and either fitting a simple functional form (e.g.~a log-normal distribution, as done in the main NANOGrav analysis \cite{Nanograv_23}), or using more elaborate techniques such as normalizing flows \cite{Laal_25}. Recently, Refs.~\cite{Ellis_23, Sato-Polito_25} derived a general Fourier-transform expression for the PDF of $h_c^2$ given a SMBHB model. While manageable for analytic population models, this expression is not readily applicable to simulation-based SMBHB models typically used for PTA data interpretation. In an appendix, Ref.~\cite{Ellis_23} proved that, in the large-source-count limit, the PDF of $h_c^2$ sourced by a population of circular binaries asymptotes to a universal self-similar limit, which they provided in an integral form. They showed that the width of this asymptotic PDF is inversely proportional to the 1/3 power of the average number of sources in a given band. 

In this work, we expand on the large-source-count result of Ref.~\cite{Ellis_23} in several significant ways, and cast it in a form that is transparent and readily usable for practical applications. First and foremost, we provide a simple closed-form analytic expression for the PDF of $h_c^2$ in the large-source-count limit. Indeed, we prove that the integral form first derived in the appendix of Ref.~\cite{Ellis_23} is the maximally-skewed stable distribution with index $3/2$ known as the map-Airy distribution \cite{Banderier_01}, suitably shifted and rescaled. Second, we clearly identify the \emph{effective} number of in-band sources $N_k$ as the large expansion parameter. Concretely, we define $N_k$ in terms of the band-averages of the mean characteristic strain squared $\overline{h_c^2}(f)$ and of the \emph{shot-noise strain scale} $\overline{h_0^3}(f)$, a new summary statistic of the SMBHB population, which only depends on its local (redshift $z \ll 1$) properties. Third, we show explicitly that this asymptotic large-source-count PDF holds for arbitrary SMBHB populations, regardless of the orbital hardening mechanism, and, importantly, for both circular and eccentric binaries. We give an explicit expression for the band-averaged shot-noise strain scale $\overline{h_{0, k}^3}$ for eccentric binaries, emphasizing subtleties arising from the fact that they emit GW radiation at multiple frequencies. In passing, we uncover a subtle conceptual flaw in the way the realization-dependent characteristic strain squared is usually computed. We point out that one should account for the full inclination dependence of the GW energy flux emitted by each binary in any given realization, rather than average over it as is usually done. Even though this flux varies by a factor 8 from edge-on to face-on circular binaries, it turns out that properly including this effect only broaden the distribution of $h_c^2$ by 10\% in the large-source-count regime.

In addition, we discuss the practical implications of our results for the modeling of discreteness effects from SMBHBs in PTA data analysis. Using a simple but physically motivated SMBHB population model, we show that the effective number of sources $N_k$ is indeed expected to be large at the lowest frequency bands probed by PTAs, where their sensitivity is highest. We then demonstrate that the large-source-count limit approximates the exact PDF of $h_c^2$ with an accuracy comparable to that of cutting-edge machine-learning-based approaches \cite{Laal_25}, and significantly better than the log-normal approximation used by the NANOGrav collaboration \cite{Nanograv_23}. Lastly, we argue that a large effective number of sources is a necessary condition for the commonly used Gaussian mixture approximation to be valid. These considerations make a strong case for using the simple large-source-count analytic expression derived in this work when analyzing PTA data within the Gaussian-mixture approximation. We spell out explicitly how our results can be incorporated into existing simulation and analysis pipelines.

The rest of this paper is organized as follows. In Sec.~\ref{sec:GWB}, we recall general expressions for the GWB from a population of circular binaries and write the exact integral expression for the PDF of $h_{c}^2$, including the full inclination dependence. In Sec.~\ref{sec:large-Nk}, we re-derive the universal large-source-count limit for circular binaries, which we generalize to eccentric binaries in Sec.~\ref{sec:eccentricity}. We compare the large-source-count PDF against exact numerical calculations for a simple but realistic SMBHB population model in Sec.~\ref{sec:model}, and discuss applications to PTA data analysis in Sec.~\ref{sec:PTA}. We conclude in Sec.~\ref{sec:conclusion}. The mathematical proof of the equivalence of the integral form of the large-source-count PDF with the map-Airy distribution is given in the Appendix.

\section{Gravitational-wave background from circular binaries}\label{sec:GWB} 

\subsection{Conventions}

Throughout this paper we use geometric units ($G = c = 1$). All frequencies $f$ are positive.

We define the characteristic strain squared $h_c^2(f)$ in the usual way as the dimensionless scalar power spectrum of the GW strain (see e.g.~Ref.~\cite{YAH_20} for precise definitions). 
%$h_{ij}(t)$, i.e.~such that
%\be
%\langle h_{ij}(t) h_{ij}(t + \tau) \rangle_t = \int \frac{d f}{f} h_c^2(f) \cos(f \tau),
%\ee
%where the average is over a time longer than the coherence time. 
It is related to the GW energy density per frequency interval through \cite{Phinney_01} 
\be
h_c^2(f) \equiv \frac4{\pi f^2} \frac{d \rho_{\rm GW}}{d \ln f} = \frac{4}{\pi f} \frac{d \rho_{\rm GW}}{d f}. \label{eq:hc2-def}
\ee
This equation can be taken as a definition of $h_c^2(f)$. 

In practice we will deal with finite frequency bands $B_k \equiv [f_k - \Delta f/2, f_k + \Delta f/2]$ centered at frequencies $f_k$ with width $\Delta f$, and will denote the band-averaged characteristic strain squared by
\be
h_{c, k}^2 \equiv \frac{f_k}{\Delta f} \int_{f_k - \Delta f/2}^{f_k + \Delta f/2} \frac{df}{f} h_c^2(f) \equiv \frac{f_k}{\Delta f} \int_{B_k} \frac{df}{f} h_c^2(f), \label{eq:hk-def}
\ee
where from here on $\int_{B_k}$ is a shorthand for the integral over the $k$-th frequency band.

Throughout the paper, we will refer to the PDFs of various quantities $X$, and denote these functions by the same letter $P(X)$. The argument of $P(X)$ should make its meaning unambiguous, and PDFs are always normalized such that $\int dX~ P(X) = 1$. 

\subsection{Contribution from a single binary}

Consider a circular binary with chirp mass $\mathcal{M}$, rest-frame GW frequency $f_r$ (which is twice its orbital frequency), inclination $\iota \in [0, \pi/2)$ of the orbital plane with respect to the plane of the sky, and redshift $z$ corresponding to radial comoving distance $r$ and luminosity distance $d_L = (1 + z) r$. The GW flux -- hence energy density, in geometric units -- contributed by this binary at Earth per frequency interval is 
\barr
\frac{d \rho_{\rm GW}}{d f} &=& \frac{32}{5} \frac{(\mathcal{M} \pi f_r)^{10/3}}{4 \pi d_L^2} g(\iota) ~\delta(f - f_r/(1+z)),  \label{eq:drho/dlnf}
\earr
where the inclination dependence is given by \cite{Peters_63}
\be
g(\iota) = \frac{5}{16} (1 + 6 \cos^2 \iota + \cos^4 \iota).
\ee
This factor accounts for the fact that the flux of GW energy emitted by the binary varies with inclination, by as much as a factor 8 between edge-on ($\iota = \pi/2$) and face-on $(\iota = 0)$ orientations. The factor $g(\iota)$ averages to unity over inclinations,  $\langle g(\iota)\rangle \equiv \int_0^1 d \cos \iota ~g(\iota) =  1$.

Using Eq.~\eqref{eq:hc2-def} and denoting by $f_0 \equiv f_r/(1 + z)$ the redshifted source rest-frame frequency, the characteristic strain squared contributed by the source is
\barr
h_c^2(f) &=& \mathfrak{h}^2(\mathcal{M}, r, f_0)~ g(\iota)~ f \delta(f - f_0),\\
\mathfrak{h}(\mathcal{M}, r, f_0) &\equiv& \sqrt{\frac{32}{5}} (1 + z)^{2/3} \frac{\mathcal{M}^{5/3} (\pi f_0)^{2/3}}{r}, \label{eq:h-def} 
\earr
where the redshift $z$ is uniquely defined from the radial comoving distance $r$. The band-averaged characteristic strain squared contributed by a single source is then
\be
h_{c, k}^2 = \frac{f_k}{\Delta f} \mathfrak{h}^2(\mathcal{M}, r, f_0)~ g(\iota)~\mathbf{1}_{f_0 \in B_k}, \label{eq:hk2-sum}
\ee
where the function $\mathbf{1}_X$ is unity if its subscript is true and vanishes otherwise. 

\subsection{Mean $h_{c, k}^2$ from a statistically-isotropic collection of sources}

We now consider a collection of binaries with parameters $\mathcal{M}_b, r_b, f_{0b}, \iota_b$.  The total band-averaged characteristic strain squared is obtained from summing the contributions from all in-band binaries: 
\be
h_{c, k}^2 = \frac{f_k}{\Delta f} \sum_b \mathfrak{h}^2(\mathcal{M}_b, r_b, f_{0 b}) g(\iota_b) \mathbf{1}_{f_{0b} \in B_k}. \label{eq:hci total}
\ee
We assume that the binaries are drawn from a distribution which is statistically isotropic in the sky and uniform in inclination, with mean number density in the 6-dimensional source parameter space
\barr
\frac{d N}{d^6 \Lambda} &\equiv& \frac{d N}{d^3 r d \mathcal{M} d f_0 d \cos \iota }= \frac{d n(r)}{d \mathcal{M} d f_0} = \frac1{f_0} \frac{d n(r)}{d \mathcal{M} d \ln f_0},~
\earr
where $d n(r)/d \mathcal{M}d f_0$ is the mean comoving number density of sources per chirp mass and redshifted rest-frame frequency interval. The mean band-averaged characteristic strain squared is then
\barr
\overline{h_{c, k}^2} &=& \frac{f_k}{\Delta f} \int d^6 \Lambda \frac{d N}{d^6 \Lambda} \mathfrak{h}^2(\mathcal{M}, r, f_0) g(\iota)  \mathbf{1}_{f_0 \in B_k}, \label{eq:mean-hk2}\\
&=& \frac{f_k}{\Delta f} \int_{B_k} \frac{df_0}{f_0}~ \overline{h_c^2}(f_0),
\earr
where the \emph{mean} characteristic strain squared is given by
\barr
\overline{h_c^2}(f) \equiv \int 4\pi r^2 dr ~d \mathcal{M}~ \frac{d n(r)}{d \mathcal{M} d \ln f} ~ \mathfrak{h}^2(\mathcal{M}, r, f),~~~\label{eq:mean-hc2}
\earr
and we used $\langle g(\iota)\rangle = 1$. This is equivalent to the expression given in Ref.~\cite{Sesana_08}, generalizing Phinney's well-known result \cite{Phinney_01} to a general SMBHB number density $dn/d \mathcal{M}d \ln f_0$.

\subsection{PDF of the band-averaged characteristic strain squared}

The main quantity of interest in this work is the PDF $P(h_{c, k}^2)$ of the band-average characteristic strain squared.  Assuming the actual number of binaries in any finite bin in their 6-dimensional parameter space is Poisson distributed, it is straightforward to show that \cite{Ellis_23, Sato-Polito_25}
\be
P(h_{c, k}^2)  =\int \frac{d u}{2 \pi} \exp\left[- i u h_{c, k}^2 + K(u)\right],  \label{eq:PDF}\\
\ee
where the cumulant generating function (CGF) is 
\barr
K(u) &\equiv& \int d^6 \Lambda \frac{d N}{d^6 \Lambda} \nonumber\\
&& \times  \left(\exp\left[i u \frac{f_k}{\Delta f} \mathfrak{h}^2(\mathcal{M}, r, f_0) g(\iota)\mathbf{1}_{f_0 \in B_k}\right] -1 \right)~~~~~\label{eq:K(u)-first}\\
&=& \int 4 \pi r^2 dr ~d \mathcal{M}~ d f_0 ~ d \cos \iota ~\frac{d n(r)}{d \mathcal{M} d f_0}  \nonumber\\
&& \times \left(\exp\left[i u \frac{f_k}{\Delta f} \mathfrak{h}^2(\mathcal{M}, r, f_0) g(\iota)\mathbf{1}_{f_0 \in B_k} \right] -1 \right).~~~~~ \label{eq:K(u)}
\earr
It is straightforward to show that this PDF indeed has mean $\overline{h_{c, k}^2}$, given by Eq.~\eqref{eq:mean-hk2}. However, unless the source density vanishes fast enough towards the origin, such that $\frac1{r} d n(r)/d \mathcal{M}d f_0 \rightarrow 0$ as $r \rightarrow 0$, the variance of $h_{c, k}^2$ diverges \cite{Xue_25}. This also implies that the angular power spectrum of GWB anisotropies is formally divergent \cite{Lin_26}. It is customary in the literature to impose a low-redshift cutoff (e.g.~\cite{Sato-Polito_24}), or assume a merger rate density that vanishes fast enough at the origin \cite{Lin_26}. Both approximations are difficult to justify\footnote{A better-motivated cutoff would be based on the non-detection of individual SMBHBs, but to our knowledge such sensitivity-based cutoffs are not currently implemented.}. All this means is that a moment expansion is not a useful representation of the PDF of $h_{c, k}^2$, which should simply be computed as a whole.

Before closing this section, let us note that nearly identical versions of this PDF were first given in Refs.~\cite{Ellis_23, Sato-Polito_25, Xue_25}, with the substitution $g(\iota) \rightarrow 1$. This is because these works computed the PDF of a different quantity, defined as in Eq.~\eqref{eq:hci total} but with $g(\iota_b) \rightarrow 1$. While one is free to define such a quantity \emph{mathematically} (in the spirit of Ref.~\cite{Xue_25}), it is important to understand that it \emph{does not have physical meaning}. Specifically, it does \emph{not} represent a realization-dependent GW energy density (or characteristic strain squared), as the latter is proportional to the sum of \emph{inclination-dependent} fluxes of all sources.

The main purpose of this paper is to provide a simple self-similar form for the PDF $P(h_{c, k}^2)$ valid in the large-source-number limit, which we now define in detail.

\section{The large-source-count limit} \label{sec:large-Nk}

\subsection{Universal scaling of the large-strain source distribution} \label{sec:large-S}

We define the number of sources per interval of contributed in-band characteristic strain squared  as follows:
\barr
\frac{d N}{d S} &\equiv& \int d^3r ~d \mathcal{M}~ d f_0~ d \cos \iota~ \frac{d n(r)}{d \mathcal{M} d f_0} \nonumber \\
&& \times \delta\left(S - \frac{f_k}{\Delta f} \mathfrak{h}^2(\mathcal{M}, r, f_0)g(\iota) \mathbf{1}_{f_0 \in B_k}\right). \label{eq:dNdh2}
\earr
We will sometimes refer to it as the ``source flux distribution", since $\mathfrak{h}^2g(\iota)$ is proportional to the GW energy flux from a given source. We can eliminate the Dirac delta function in Eq.~\eqref{eq:dNdh2} by integrating over $r$. Changing variables, we rewrite the delta function as 
\barr
\delta\left(S - \frac{f_k}{\Delta f} \mathfrak{h}^2 g \mathbf{1}_{f_0}\right) &=& \frac{R_\Lambda}{2 S^{3/2}} \frac{(1 + z_S)^{2/3}}{J(r_S)} \delta(r- r_S), \label{eq:dirac-new var}\\
J(r) &\equiv& 1 - \frac23 \frac{r}{1 + z} \frac{\partial z}{\partial r},~~~~\label{eq:Jacobian}
\earr
where the distance $r_S$ is the implicit solution\footnote{In reality there are exactly 2 solutions to Eq.~\eqref{eq:r-equation} if the right-hand-side is less than $0.58/ H_0$ (and none otherwise), one of them with $z < z_*$ and the other with $z > z_*$, where $z_* \approx 2.62$ is the redshift at which the Jacobian \eqref{eq:Jacobian} vanishes. For any physically relevant model, the contributions to the GWB from $z \gtrsim z_*$ is negligible, and we may therefore safely keep only the low-redshift solution. Numbers quoted here assume the Planck 2018 \cite{Planck_20} best-fit value of the matter density parameter $\Omega_m$. \label{foot:sol}} of
\be
\frac{r_S}{(1+z_S)^{2/3}} =\frac{R_\Lambda}{\sqrt{S}}, \label{eq:r-equation}
\ee
with $z_S \equiv z(r_S)$, and the lengthscale $R_\Lambda$ is defined as 
\be
R_\Lambda \equiv \left(\frac{32}{5} \frac{f_k}{\Delta f}  g(\iota)\right)^{1/2} \mathcal{M}^{5/3}( \pi f_0)^{2/3}~ \mathbf{1}_{f_0 \in B_k}, \label{eq:R_Lambda}
\ee
where we used $(\mathbf{1}_X)^\alpha = \mathbf{1}_X$ for any $\alpha > 0$. We may then rewrite $dN/dS$ in the somewhat more explicit form:
\barr
\frac{d N}{dS} &=& \frac{2 \pi}{S^{5/2}} \left( \frac{32}{5} \frac{f_k}{\Delta f} \right)^{3/2} \int_{f_0 \in B_k} d \mathcal{M}~ d f_0 ~d \cos \iota \nonumber\\
&& \times  \mathcal{M}^5 (\pi f_0)^2 ~ g(\iota)^{3/2}~ \frac{(1+z_S)^2}{J (r_S)}~ \frac{d n(r_S)}{d \mathcal{M} d f_0} \nonumber\\
&=& \frac{2 \pi}{S^{5/2}} \left(\frac{f_k}{\Delta f} \right)^{3/2} \int_{f_0 \in B_k} d \mathcal{M}~ d f_0 ~d \cos \iota~ \nonumber\\
&& \times \frac{1}{J(r_S)} \frac{d n(r_S)}{d \mathcal{M} d f_0}~ r_S^3 ~\mathfrak{h}^3(\mathcal{M}, f_0, r_S) ~g(\iota)^{3/2}. \label{eq:dN-dS explicit}
 \earr
For sufficiently large $S$, for any physically relevant $\mathcal{M}$ and $f_0$, and any inclination angle $\iota$, the right-hand-side of Eq.~\eqref{eq:r-equation} is much smaller than the Hubble radius, and so is the physically relevant solution $r_S$ (see footnote \ref{foot:sol}), with corresponding redshift $z_S \ll 1$. In that limit, $J(r_S) \approx 1$ and we can evaluate $dn(r_S)/d \mathcal{M}df_0$ at $r  = 0$. We then obtain the following large-strain scaling for $d N/dS$ 
\be
\frac{d N}{dS} \approx \frac1{4 \sqrt{\pi}} \sqrt{\frac{f_k}{\Delta f}}~ \frac{\overline{h_{0, k}^3}}{S^{5/2}}, \label{eq:dN-dS-large-S}
\ee
where $\overline{h_{0, k}^3} \equiv (f_k/\Delta f) \int_{B_k} d\ln f_0~ \overline{h_0^3}(f_0)$ is the band-average of the (cubic) \emph{shot-noise strain scale}, which is defined from the \emph{local} ($z \ll 1 $) SMBHB population as
\begin{empheq}[box=\fbox]{equation}
\begin{aligned}
\overline{h_0^3}(f)  &=  (4\pi)^{3/2} \langle g(\iota)^{3/2} \rangle \\
& \times \lim_{r \rightarrow 0}\left[\int d \mathcal{M} ~\frac{d n(r)}{d \mathcal{M} d \ln f} ~ r^3~ \mathfrak{h}^3(\mathcal{M}, f, r)\right]. 
\end{aligned}
\label{eq:h0^3-def}
\end{empheq}
As we will see more sharply later on, $\overline{h_0^3}(f)$ characterizes the width of the PDF of the GWB, and is thus complementary to the mean characteristic strain squared $\overline{h_c^2}(f)$. Note its precise definition \eqref{eq:h0^3-def}, which does \emph{not} involve a volume integral of $\mathfrak{h}^3$, which would diverge at the origin. The specific normalization convention is chosen so that subsequent expressions take a compact form.

The large-flux scaling $dN/dS \propto S^{-5/2}$ was pointed out in Ref.~\cite{Ellis_23}, and its universality emphasized recently in Ref.~\cite{Raidal_26b}. Note that there is no universal low-flux scaling for the source density, as it depends on the low-mass dependence of the source density $d n/d \mathcal{M}d f_0$ \cite{Sato-Polito_25}. However, in order to give a finite mean GWB density, it must be that $S^2 (d N/dS) \rightarrow 0$ as $S \rightarrow 0$. 

\subsection{Characteristic source number}

We now define an effective number of sources contributing to the GWB similarly to Ref.~\cite{Sato-Polito_25}. In terms of the source flux distribution $dN/dS$, the mean band-averaged characteristic strain squared is simply
\be
\overline{h_{c, k}^2} =\int dS~ \frac{d N}{d S}~  S = \int d \ln S~ \frac{d N}{dS}~ S^2.
\ee
Since this integral is convergent, and the integrand $(dN/dS) S^2$ falls off as $S^{-1/2}$ at large $S$, the latter must peak at some characteristic flux $S_*$, implying that $\overline{h_{c, k}^2} \sim S_*^2 (d N/dS)_*$. Extrapolating the large-flux behavior Eq.~\eqref{eq:dN-dS-large-S} down to $S_*$, we moreover have $(d N/dS)_* \sim \sqrt{f_k/\Delta f}~\overline{h_{0, k}^3}/S_*^{5/2}$, implying $\overline{h_{c, k}^2} \sim \sqrt{f_k/\Delta f}~\overline{h_{0, k}^3}/ S_*^{1/2}$, hence $S_* \sim (f_k/\Delta f) (\overline{h_{0, k}^3})^2/(\overline{h_{c, k}^2})^2$. The characteristic number of sources contributing to the GWB in the $k$-th band is then $N_k \sim \overline{h_{c, k}^2}/S_* \sim (\Delta f/f_k)(\overline{h_{c, k}^2})^3/(\overline{h_{0, k}^3})^2$.

These considerations bring us to \emph{define} the following dimensionless parameter as a characteristic number of sources in a given frequency band:
\be 
\boxed{N_k \equiv \frac{\Delta f}{f_k} \frac{\left(\overline{h_{c, k}^2}\right)^3} {\left(\overline{h_{0, k}^3}\right)^2}}~. \label{eq:N_k-def}
\ee
In the narrow-band limit, $\overline{h_{c, k}^2} \approx \overline{h_c^2}(f_k)$ and $\overline{h_{0, k}^3} \approx \overline{h_0^3}(f_k)$, implying that the effective number of in-band sources is linear in the band width $\Delta f$, as one would expect intuitively.

\subsection{PDF of $h_{c, k}^2$ in the large-source-count limit}

We now derive a self-similar, universal form for the PDF of $h_{c, k}^2$ in the large-effective-source-count limit $N_k \gg 1$ (see also the Appendix of Ref.~\cite{Ellis_23}).

We define the rescaled source flux distribution $d N/ds \equiv  \overline{h_{c, k}^2} (d N/dS)$, where $s \equiv S/\overline{h_{c, k}^2}$. The large-flux asymptote \eqref{eq:dN-dS-large-S} holds for $S \gg S_* \sim \overline{h_{c, k}^2}/N_k$, or equivalently $s \gg 1/N_k$. We thus write, in all generality,
\be
\frac{dN}{ds}= \frac1{4 \sqrt{\pi}} \frac1{\sqrt{N_k}} \frac{\mathcal{F}(s N_k)}{s^{5/2}},  \label{eq:dNds-F}
\ee
where we used the definition \eqref{eq:N_k-def} of $N_k$, and $\mathcal{F}$ is some function which depends on the specific SMBHB model, but that universally has the limit  
\be
\mathcal{F}(x \gg 1) = 1. \label{eq:mathcalF-lim}
\ee
We then rewrite Eq.~\eqref{eq:PDF} to obtain the PDF of the dimensionless variable $y \equiv h_{c, k}^2/ \overline{h_{c, k}^2}$ as follows
\barr
P(y) &=& \int_{-\infty}^{\infty} \frac{d v}{2 \pi} \exp\left[i v (1-y) + L(v)\right], \label{eq:P(rho)-2}\\
L(v) &\equiv& \int_0^\infty ds \frac{d N}{d s} (e^{i v s } -1 - i v s),\label{eq:L(v)}
\earr
where we have used the fact that $\int ds \frac{dN}{ds} s = 1$ by construction. Since $L(-v) = L(v)^*$, we may focus on $v > 0$. We insert the general form \eqref{eq:dNds-F} into Eq.~\eqref{eq:L(v)}, and change variables to $t = v s$, obtaining
\barr
L(v) &=& \frac{v^{3/2}}{\sqrt{N_k}} \mathcal{J}(v/N_k), \\
\mathcal{J}(\tau) &\equiv& \frac1{4 \sqrt{\pi}} \int_0^{\infty} dt~ \mathcal{F}(t/\tau)\frac{e^{i t} - 1 - i t}{t^{5/2}}. \label{eq:mathcal(J)}
\earr
In the limit that $N_k  \gg v$, using the universal limit $\mathcal{F}(x \gg 1)= 1$, we may replace $\mathcal{J}(v/N_k)$ with its limit 
\be
\mathcal{J}(0) = \frac1{4 \sqrt{\pi}} \int_0^{\infty} dt~ \frac{e^{i t} - 1 - i t}{t^{5/2}} = - \frac{1 + i}{3 \sqrt{2}}, 
\ee
so that
\barr
L(v \ll N_k) = - \frac{1 + i}{3 \sqrt{2}} \frac{v^{3/2}}{\sqrt{N_k}}. \label{eq:L(v)-approx}
\earr
Inserting this expression into Eq.~\eqref{eq:P(rho)-2}, and using $L(-v) = L(v)^*$, we find that the PDF of $y = h_{c, k}^2/\overline{h_{c, k}^2}$ is approximately
\barr
P(y) \simeq \int_0^{\infty} \frac{d v}{\pi} \cos\left((y -1) v +  \frac{v^{3/2}}{3\sqrt{2N_k}}  \right)\nonumber\\
\times \exp\left(- \frac{v^{3/2}}{3\sqrt{2N_k}} \right).~~~~~
\earr
Finally, changing variables to $t \equiv v/N_k^{1/3}$, we arrive at the following self-similar form of the large-$N_k$ PDF of the band-averaged characteristic strain squared:
\be
\boxed{P(h_{c, k}^2) \simeq \frac{N_k^{1/3}}{\overline{h_{c, k}^2}} \mathcal{P}\left(N_k^{1/3} \left(h_{c, k}^2/\overline{h_{c, k}^2} -1\right)\right)},\label{eq:main-1}
\ee
or, equivalently, using Eq.~\eqref{eq:N_k-def},
\be
P(h_{c, k}^2) \simeq \frac{(\Delta f/f_k)^{1/3}}{\left(\overline{h_{0, k}^3}\right)^{2/3}} \mathcal{P}\left((\Delta f/f_k)^{1/3}\frac{h_{c, k}^2 - \overline{h_{c, k}^2}}{\left(\overline{h_{0, k}^3}\right)^{2/3}}\right), \label{eq:main-2}
\ee
where the universal probability distribution $\mathcal{P}(x)$ is defined through its Fourier transform
\barr
\mathcal{P}(x) \equiv \int_0^{\infty} \frac{dt}{\pi}~ \cos\left(x t + \frac{t^{3/2}}{3 \sqrt{2}} \right) e^{-  \frac{ t^{3/2}}{3 \sqrt{2}}}.~~~~ \label{eq:P-integral}
\earr
Up to parameter redefinitions, Eq.~\eqref{eq:main-2} is identical to Eq.~(A.15) of Ref.~\cite{Ellis_23}, and up to a rescaling and contour rotation, Eq.~\eqref{eq:P-integral} is identical to their Eq.~(A.16).

The PDF given in Eq.~\eqref{eq:P-integral} belongs to the general class of \emph{stable distributions}: it satisfies the property that a linear combination of variables with this distribution still has the same distribution, up to shifting and rescaling. In the Appendix, we prove that, up to a reflection about the origin, Eq.~\eqref{eq:P-integral} is the maximally-skewed stable distribution with stability parameter 3/2 known as the \emph{map-Airy distribution} \cite{Banderier_01, Weisstein_MapAiry}. It can be expressed explicitly in terms of the Airy function and its derivative:
\be
\boxed{\mathcal{P}(x) = - 2 e^{2x^3/3}\left(x \textrm{Ai}(x^2) + \textrm{Ai}'(x^2) \right)}~. \label{eq:Airy-map}
\ee

Equation \eqref{eq:main-1} or its equivalent \eqref{eq:main-2}, along with Eq.~\eqref{eq:Airy-map}, and the definition of the characteristic number of in-band sources Eqs.~\eqref{eq:h0^3-def} and \eqref{eq:N_k-def}, constitute our main result: they provide a simple and universal form of the probability distribution of the GWB sourced by inspiralling circular binaries in the large-source-count limit, parametrized by the mean characteristic strain squared $\overline{h_{c, k}^2}$ and the effective in-band source number $N_{k}$. In the limit that $N_k \rightarrow \infty$, we recover $P(h_{c, k}^2) \rightarrow \delta(h_{c, k}^2 - \overline{h_{c, k}^2})$. For a large but finite source count, the PDF has relative width of order $N_k^{-1/3}$. We see, from Eq.~\eqref{eq:main-2}, that the shot-noise strain scale $\overline{h_0^3}(f)$ introduced in Eq.~\eqref{eq:h0^3-def} is a direct measure of the width of the distribution of $h_{c, k}^2$ about its mean, in the large-$N_k$ limit. While the full redshift dependence of the SMBHB population determines the mean $\overline{h_{c, k}^2}$, in the large-$N_k$ limit it is only its local ($z \ll 1$) properties that determine the spread of $h_{c, k}^2$ about its mean, as this spread is determined by the loud nearby sources. In passing, we also see from Eqs.~\eqref{eq:h0^3-def} and \eqref{eq:main-2} that properly including the inclination dependence widens the PDF by a factor $\langle g(\iota)^{3/2} \rangle^{2/3} \approx 1.096$, i.e.~has a $\sim 10\%$ effect. 

\begin{figure}
\includegraphics[width = \linewidth]{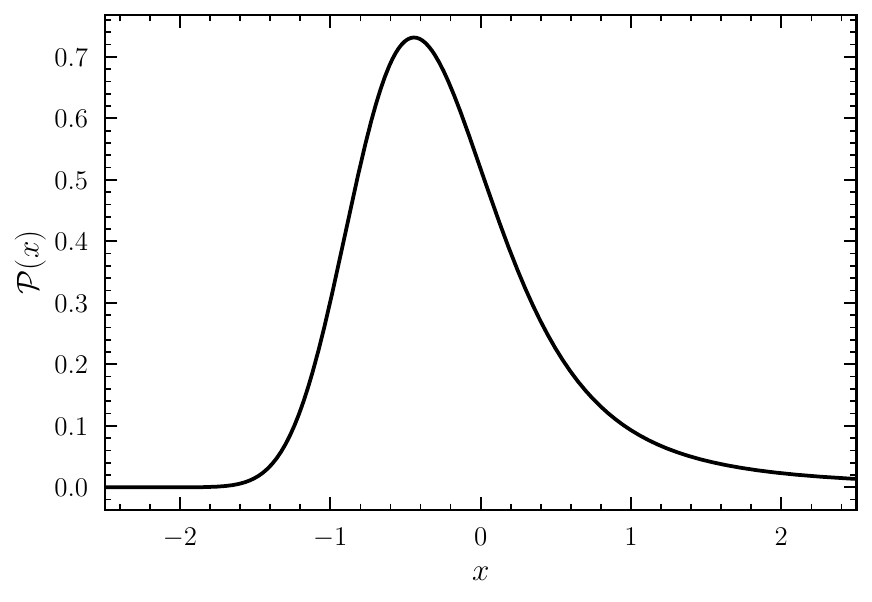}
\caption{Universal shape of the large-source-count probability distribution for the GWB background, which is the (reflected) map-Airy distribution \citep{Banderier_01}, given explicitly in Eq.~\eqref{eq:Airy-map}.} \label{fig:P(x)}
\end{figure}

We show the universal probability distribution $\mathcal{P}(x)$ in Fig.~\ref{fig:P(x)}. It has zero mean, median $x_{\rm med} \approx -0.273$, peaks at $x_{\rm peak} \approx - 0.443$ and has full width at half maximum FWHM $\approx 1.167$. As we review in more detail below, it scales as $\sim 1/x^{5/2}$ at large positive $x$ and decays exponentially at large negative $x$ \cite{Ellis_23}. 

\begin{figure*}[ht]
\includegraphics[width = \linewidth]{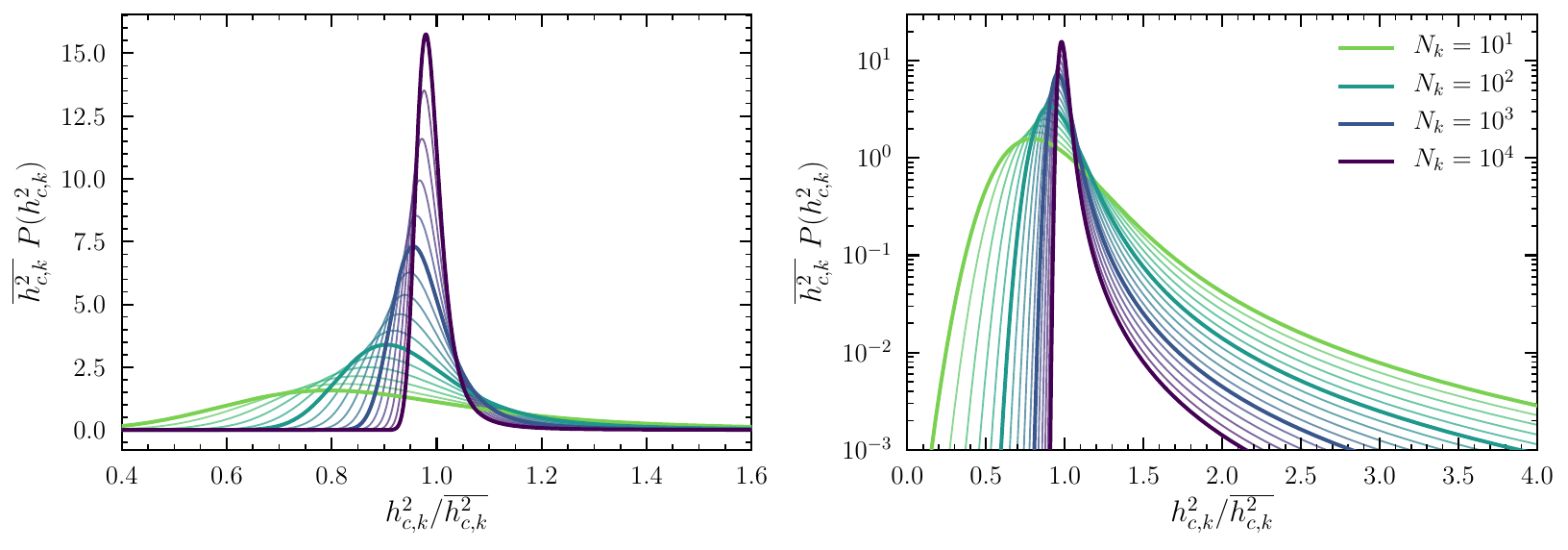}
\caption{Probability distribution of the band-averaged GWB characteristic strain squared in the large-source-count limit, given by Eq.~\eqref{eq:main-1}. The left panel highlights the behavior of $P(h_{c, k}^2)$ near the mean, while the right panel shows a broader region highlighting the large-$h_{c, k}^2$ power-law asymptotic behavior, and the exponential cutoff at small $h_{c, k}^2$. The values of $N_k$ are logarithmically distributed between $10$ and $10^4$, with a few specific values highlighted by thicker lines.}\label{fig:P(rho)}
\end{figure*}

We also show the self-similar large-$N_k$ PDF $P(h_{c, k}^2)$ for several illustrative values of $N_k$ in Fig.~\ref{fig:P(rho)}.

\subsection{Asymptotic behavior}

The universal PDF $\mathcal{P}(x)$ has the following asymptotic limits (see also Ref.~\cite{Ellis_23}):
\barr
\mathcal{P}(x \gg 1) &\approx&  \frac1{4 \sqrt{\pi}~ x^{5/2}}, \\
\mathcal{P}(x \ll - 1) &\approx& \frac{2|x|^{1/2}}{\sqrt{\pi}}e^{- 4 |x|^3/3}. \label{eq:small-rho}
\earr
The former limit implies that, for $h_{c, k}^2/\overline{h_{c, k}^2}-1 \gg 1/N_k^{1/3}$, 
\barr
P(h_{c, k}^2) \approx \frac{d N}{dS}\Big{|}_{S = h_{c, k}^2 - \overline{h_{c, k}^2}}~,
\earr
where we substituted $N_k$ with its definition Eq.~\eqref{eq:N_k-def} and used the asymptotic form \eqref{eq:dN-dS-large-S}. This known large-$h_{c, k}^2$ scaling \cite{Ellis_23, Ellis_24} has a simple physical meaning: the large-$h_{c, k}^2$ limit is dominated by the rare nearby single sources that add a flux $(h_{c, k}^2 - \overline{h_{c, k}^2})$ on top of the mean $\overline{h_{c, k}^2}$ contributed by the bulk of ``typical", distant sources. 

On the other hand, the small-$h_{c, k}^2$ limit corresponds to a collective downward fluctuation of a large number of ``typical" sources, and we thus anticipate from Poisson statistics that in that regime $P(h_{c, k}^2)$ should scale as $\sim e^{- N_k F(h_{c, k}^2/\overline{h_{c, k}^2})}$, where $F$ is some function that depends on the shape of $d N/dS$ near its peak. This is indeed the scaling that one obtains from Eq.~\eqref{eq:small-rho}. However, since $\mathcal{P}(x)$ is obtained from the large-flux asymptotic behavior of $d N/dS$, and does not use any information about its shape near its peak, we expect the asymptotic behavior predicted by our approximation for $h_{c, k}^2/\overline{h_{c, k}^2} -1 \ll -1/N_k^{1/3}$ to be a poor representation of the correct small-$h_{c, k}^2$ behavior of $P(h_{c, k}^2)$. Another way to see this is from the mathematical proof of Eq.~\eqref{eq:main-1}: in the Fourier domain, the asymptotic form is reached for $v \ll N_k$, and the approximation Eq.~\eqref{eq:main-1} thus neglects the contributions of Fourier modes with $v \gtrsim N_k$. It may therefore not accurately capture very sharp transitions in $h_{c, k}^2/\overline{h_{c, k}^2}$, such as the exponential drop at small $h_{c, k}^2/\overline{h_{c, k}^2}$. Given that $P(h_{c, k}^2)$ is exponentially suppressed in that regime anyway, in practice this inaccuracy should not be critical. We will quantify these points in Section \ref{sec:model}, where we compute the exact $P(h_{c, k}^2)$ for a specific SMBHB model.

\section{Generalization to eccentric binaries}\label{sec:eccentricity}

It is possible that the eccentricity of SMBHBs has not fully dissipated by the time they enter the nanohertz band, which could have significant effects on the GWB spectrum \cite{Kelley_17, Raidal_24}. Ultimately, the large-$N$ limit we derived above relies on the inverse-square-distance scaling of the gravitational-wave energy flux, which translates into a universal large-flux asymptotic scaling of the source number distribution. This scaling is general and independent of eccentricity, and the asymptotic limit we derived therefore carries over to a population of eccentric SMBHBs. In this section we generalize the calculation to eccentric binaries, which mostly parallels the circular-binary case, with some important qualitative differences that we highlight.

\subsection{Contribution from a single binary}

Consider a binary with chirp mass $\mathcal{M}$, comoving distance $r$, eccentricity $e$, and redshifted rest-frame orbital frequency $f_{\rm orb}/(1 + z) = f_0/2$, such that $f_0$ keeps the same meaning as for circular binaries. The orientation of its orbit with respect to the plane of the sky is characterized by two angles: the inclination $\iota$, and the argument of pericenter $\omega$ (the angle between the intersection of the orbital and sky planes, and the direction of the pericenter). Such sources radiate gravitational-wave power at multiple integers of the orbital frequency \cite{Peters_63}. The characteristic strain squared (again defined through Eq.~\eqref{eq:hc2-def} from the GW energy density or energy flux) contributed by such a binary is then
\be
h_c^2(f) = \mathfrak{h}^2(\mathcal{M}, r, f_0) \sum_{n = 1}^{\infty} g_n(e, \iota, \omega)~ f \delta(f - n f_0/2),
\ee  
and the band-averaged characteristic strain squared then becomes
\be
h_{c, k}^2 = \frac{f_k}{\Delta f} \mathfrak{h}^2(\mathcal{M}, r, f_0) \sum_{n =1}^{\infty} g_n(e, \iota, \omega) ~\mathbf{1}_{n f_0/2 ~\in B_k},
\ee
where $\mathfrak{h}$ was defined in Eq.~\eqref{eq:h-def}. The angular averages of the harmonic amplitudes  
\be
g_n(e) \equiv \int_0^1 d \cos \iota \int_0^{2 \pi} \frac{d \omega}{2 \pi} g_n(e, \iota, \omega)
\ee
are well-known functions of eccentricity that can be expressed in terms of Bessel functions \cite{Peters_63}. To our knowledge, the angle-dependent coefficients $g_n(e, \iota, \omega)$ themselves are not given explicitly in the literature, but they could be straightforwardly extracted from intermediate results presented in Ref.~\cite{Peters_63}. We will not need them explicitly for the purpose of proving the theorem.

\subsection{Mean $h_{c, k}^2$ from a statistically-isotropic collection of sources}

When considering a population of randomly-oriented eccentric binaries, the binary parameter space is now 8-dimensional, as one must add $e$ and $\omega$ to the 6 parameters considered earlier. The mean number density of isotropically-distributed-and-oriented binaries in this 8-dimensional space is then
\be
\frac{d N}{d^8 \Lambda} \equiv \frac{d N}{d^3 r d \mathcal{M} de df_0 d \cos \iota d \omega} = \frac1{2\pi} \frac{d n(r)}{d \mathcal{M} de d f_0}.
\ee
The ensemble average of $h_{c, k}^2$ then becomes
\barr
\overline{h_{c, k}^2} &=& \sum_{n = 1}^{\infty} \int 4 \pi r^2 d \mathcal{M}~ d e ~ g_n(e) ~\nonumber\\
&& \times \frac{f_k}{\Delta f} \int_{n f_0/2~\in B_k} d f_0 \frac{d n(r)}{d \mathcal{M} de d f_0} \mathfrak{h}^2(\mathcal{M}, f_0, r). \label{eq:hck^2-eccentric}
\earr
Changing integration variable to (the log of) $f_n = n f_0/2$, we can factorize the integral over frequencies, and may rewrite $h_{c, k}^2$ as the band-average of the mean characteristic strain squared $\overline{h_c^2}(f)$, now defined as
\barr
\overline{h_c^2}(f) &=& \sum_{n = 1}^{\infty} \int 4 \pi r^2 d \mathcal{M}~ d e ~ g_n(e) \nonumber\\
&& \times  \frac{d n(r)}{d \mathcal{M} de d \ln f_0} \mathfrak{h}^2(\mathcal{M}, f_0, r) \Big{|}_{f_0 = 2 f/n}.
\earr

\subsection{PDF of $h_{c, k}^2$ and definition of $\overline{h_{0, k}^3}$}

The PDF of $h_{c, k}^2$ takes the form of Eq.~\eqref{eq:PDF}, where $K(u)$ is now given by the generalization of Eq.~\eqref{eq:K(u)-first} to the 8-dimensional binary parameter space, and with the substitution 
\be
g(\iota)~\mathbf{1}_{f_0 \in B_k} \longrightarrow \sum_{n \geq 1} g_n(e, \iota, \omega)~ \mathbf{1}_{n f_0/2 \in B_k}, \label{eq:substitution}
\ee
and the source flux distribution $dN/dS$ can be defined as in Eq.~\eqref{eq:dNdh2} with the same substitutions. The calculation leading to the large-flux asymptote $dN/dS$ \eqref{eq:dN-dS-large-S} is identical to the one presented in Sec.~\ref{sec:large-S}, with the substitution above, and an important difference: in Eq.~\eqref{eq:R_Lambda}, one must leave the full sum \eqref{eq:substitution} inside the square root. As a consequence, the quantity $\overline{h_{0, k}^3}$ appearing in Eq.~\eqref{eq:dN-dS-large-S} is now given by
\barr
\overline{h_{0, k}^3} \equiv (4 \pi)^{3/2} \frac{f_k}{\Delta f} \int d \mathcal{M} ~de ~df_0~  d \cos \iota~ \frac{d \omega}{2 \pi} \nonumber\\
\times \left(\sum_{n \geq 1} g_n(e, \iota, \omega) \mathbf{1}_{n f_0/2 \in B_k}\right)^{3/2} \nonumber\\
\times \lim_{r \rightarrow 0}\left[ \frac{d n(r)}{d \mathcal{M} de d f_0} r^3 \mathfrak{h}^3(\mathcal{M}, r, f_0) \right]. \label{eq:h03-eccentric}
\earr
We see that, in general, we can no longer write $\overline{h_{0, k}^3}$ as the band-average of a single, band-independent function, because of the non-trivial interference of power emitted at different harmonics, whose sum is taken to the 3/2 power. In other words, the shot-noise strain scale $\overline{h_{0, k}^3}$ is now only defined for a specific frequency band. This is in contrast with the circular case, where $\overline{h_0^3}(f)$ can be defined locally in frequency. 

With the exception of that subtlety, the derivation of the large-source-count limit follows exactly the same steps as for the case of circular binaries, leading to the same asymptotic PDF Eq.~\eqref{eq:main-1}, where the effective number of in-band sources $N_k$ is defined from $\overline{h_{0, k}^3}$ and $\overline{h_{c, k}^2}$ as given in Eq.~\eqref{eq:N_k-def}.

Notably, the fact that each eccentric binary can contribute power to multiple frequencies also implies that the PDFs of $\overline{h_{c, k}^2}$ are not independent even for non-overlapping frequency bands \cite{Raidal_24}. To account for this, one can generalize Eq.~\eqref{eq:PDF} to the joint PDF of $\overline{h_{c, k}^2}$ in multiple bands, and adequately generalize the large-source-count limit from there. We leave this more technical calculation to future work.

\section{Application to a simple SMBHB population model} \label{sec:model}

\subsection{Model definition and characteristic properties}

We now consider a simple but physically-motivated model for a population of \emph{circular} SMBHBs, for which we compute the exact $P(h_{c, k}^2)$ numerically, and compare it with the large-$N_k$ limit Eq.~\eqref{eq:main-1}. Since we have established that the full inclination dependence only broadens the PDF by a factor $\langle g(\iota)^{3/2} \rangle^{2/3} \approx 1.096$ in the large-$N_k$ limit, to simplify calculations we will set $g(\iota) \rightarrow 1$ consistently in both the exact and approximate PDFs. This means that we are truly computing the PDF of the quantity defined as Eq.~\eqref{eq:hci total} with $g(\iota_b) \rightarrow 1$. 

We assume that orbital -- hence frequency -- evolution happens on a timescale short compared to the Hubble time, so that one may accurately estimate the source number density per frequency interval with the quasi-stationary approximation
\be
\frac{d n}{d \mathcal{M} d \ln f_0} \approx \frac1{d \ln f_0/dt}~ \frac1{\mathcal{M}} \frac{d \dot{n}_{\rm mer}}{d \ln \mathcal{M}},\label{eq:qss}
\ee
where $d \dot{n}_{\rm mer}/d \ln \mathcal{M}$ is the comoving SMBHB merger rate density per logarithmic chirp mass interval. If, moreover, GW emission is the dominant orbital hardening mechanism, then 
\be
\frac{d \ln f_0}{dt} \approx \frac{d \ln f_0}{dt}\Big{|}_{\rm GW} \equiv \frac{96}{5} \mathcal{M}^{5/3} (\pi (1 + z) f_0)^{8/3}. \label{eq:df0dt-gw}
\ee
Next, we adopt a very simple model for the merger rate density, that is a slowly increasing function of redshift and approximately flat in $\log \mathcal{M}$, up to exponential cutoffs:
\be
\frac{d \dot{n}_{\rm mer}(r)}{d \ln \mathcal{M}} = \dot{n}_0 (1 + z)^2 e^{-r/r_*} e^{- (1+z)^{2/3} (\mathcal{M}/\mathcal{M}_*)^{5/3}}. \label{eq:dn_mer/dM}
\ee
This simple form for the merger rate density is similar to the class of models considered in Ref.~\cite{Middleton_16}. It emulates the main qualitative features of more elaborate frameworks, for which the SMBH mass function is obtained by convolving a log-normal distribution with a Schechter-like distribution for either the stellar mass function \cite{Nanograv_23, Liepold_24} or the stellar velocity dispersion function \cite{Sato-Polito_24b, Sato-Polito_25, Sato-Polito_25b}, and combined with a distribution of mass ratios to obtain the SMBHB chirp-mass distribution. The specific redshift prefactors and exponential dependences in Eq.~\eqref{eq:dn_mer/dM} allow for simple analytic expressions later on. Note that these choices are also physically sensible: in our model the merger rate density peaks around redshift $z_*$ (corresponding to comoving distance $r_*$), and the characteristic SMBHB mass $\mathcal{M}_*/(1 + z)^{2/5}$ decreases with redshift.

With this model, we obtain the following expression for the characteristic strain squared
\barr
\overline{h_c^2}(f) &=& \frac{4 \pi}{3} ( \pi f)^{-4/3} \int \frac{dr}{(1 + z)^{4/3}} d \mathcal{M} \mathcal{M}^{5/3} \frac{d \dot{n}_{\rm mer}}{d \mathcal{M}}~~~ \nonumber \\
&=&  \frac{4 \pi}{5} \dot{n}_0~r_*~ \mathcal{M}_*^{5/3}  ( \pi f)^{-4/3}, \label{eq:hc2-model}
\earr
where the first line is the well-known result of Ref.~\cite{Phinney_01}. The cubic shot-noise strain scale is given by 
\barr
\overline{h_0^3}(f) &=& \frac{10}{3} \left( \frac{8 \pi}{5}\right)^{3/2} (\pi f)^{-2/3}\int d \mathcal{M} \mathcal{M}^{10/3}~\frac{d \dot{n}_{\rm mer}(0)}{d \mathcal{M}} \nonumber \\
&=& 2 \left( \frac{8 \pi}{5}\right)^{3/2}  \dot{n}_0~ \mathcal{M}_*^{10/3} ~ (\pi f)^{-2/3}.
\earr

\subsection{Effective source count}

To simplify the exposition, we will make the narrow-band approximation $\int_{B_k} df~ F(f) \approx F(f_k)$, even when extrapolating our results to broad bands. This approximation does not affect any of the qualitative conclusions and can easily be lifted if desired, but it keeps expressions more compact and clearer. With this approximation, $\overline{h_{c, k}^2} \approx \overline{h_c^2}(f_k)$ and $\overline{h_{0, k}^3} \approx \overline{h_0^3}(f_k)$, and the effective number of in-band sources is then 
\barr
N_k &\approx& \frac{\pi}{32} ~ \Delta f~ \dot{n}_0 ~r_*^3 ~\mathcal{M}_*^{-5/3}  ( \pi f_k)^{-11/3}.~~~~~\label{eq:Nc-model}
\earr
%where we have defined the finite-width correction factor
%\barr
%\Lambda_k \equiv \frac1{\Delta} \frac{\left(\int_{|x - 1| < \Delta/2} \frac{dx}{x} x^{-4/3}\right)^3}{\left(\int_{|x - 1| < \Delta/2} \frac{dx}{x} x^{-2/3}\right)^2},  \ \ \ \ \Delta \equiv \frac{\Delta f}{f_k}.
%\earr
%For a narrow band with $\Delta f/f_k \ll1 $, we get $\Lambda_k \approx 1$, but for a wide band with $\Delta f/f_k =1$ (as is usually the case for lowest frequency bin analyzed by PTAs), $\Lambda_k \approx 2$. 
Using Eq.~\eqref{eq:hc2-model} to eliminate $\dot{n}_0$, we rewrite Eq.~\eqref{eq:Nc-model} as
\barr
N_k &=& \frac{5 \Delta f}{128} ~ ( \pi f_{\rm ref})^{4/3} \overline{h_c^2}(f_{\rm ref})r_*^2~ \mathcal{M}_*^{-10/3} (\pi f_k)^{-11/3},~~~
\earr
where $f_{\rm ref}$ is some reference frequency. These scalings are identical to those of Eqs.~(31) and (32) of Ref.~\cite{Sato-Polito_25}. Numerically, we find (using the Planck 2018 \cite{Planck_20} best-fit value for $H_0$)
\barr
N_k &\approx& 5.1\times 10^3 \times \frac{\Delta f}{2 ~ \textrm{nHz}} \left(\frac{h_c(f = 1/\textrm{yr})}{2.4 \times 10^{-15}} \right)^2 \nonumber\\
&& \times  \left(\frac{r_*}{2~ \textrm{Gpc}}\right)^2  \left( \frac{\mathcal{M}_*}{10^9 M_{\odot}}\right)^{-10/3} \left( \frac{f_k}{2 ~\textrm{nHz}}\right)^{-11/3} ,~~~~ \label{eq:Nc-approx}
\earr
where we have normalized $h_c \equiv \sqrt{\overline{h_c^2}}$ to the median strain amplitude reported by NANOGrav for a fiducial $f^{-2/3}$ power law \cite{nanograv15yr}, and the cutoff in comoving distance $r_*$ to 2 Gpc, corresponding to $z_* \approx 0.5$.

We thus see that, at the lowest frequencies that are best probed by NANOGrav, the effective number of sources -- as we have defined it -- is indeed expected to be large, and our approximation should thus be relevant and accurate.

It is worth emphasizing that one should not take $N_k$ too literally as a ``number of sources". First and foremost, the \emph{actual} number of sources contributing to a given band may be virtually infinite -- it is the case for our simple model, since $dn/d \ln \mathcal{M}d \ln f_0 \propto \mathcal{M}^{-5/3}$ at $\mathcal{M} \ll \mathcal{M}_*$. However, the sea of infinitely numerous faint sources only make a finite contribution to the GWB in any frequency band. The number $N_k$ is defined explicitly as an \emph{effective} number of sources, i.e.~the number of ``typical" sources (in our case, with chirp mass $\sim \mathcal{M}_*$ and at comoving distance $\sim r_*$) required to make up the bulk of the GWB. Second, the specific value of $N_k$ depends on somewhat arbitrary normalization conventions in Eqs.~\eqref{eq:h0^3-def} and \eqref{eq:N_k-def}. As a consequence, it is difficult to compare the specific numerical value we give in Eq.~\eqref{eq:Nc-approx} with numbers quoted in other references, e.g.~\cite{Nanograv_discreteness, Sato-Polito_25}. What is clear and convention-independent is the scaling with peak chirp mass and, especially, frequency, which agrees with those of Refs.~\cite{Sesana_08, Mingarelli_13, Sato-Polito_25}. Importantly, the large-source-count limit Eq.~\eqref{eq:main-1}, with $\mathcal{P}$ given by the reflected map-Airy distribution \eqref{eq:Airy-map}, holds for $N_k$ defined specifically with our convention. 

\subsection{Flux distribution}

We now compute the distribution of sources per flux interval $dN/dS$ , defined in Eq.~\eqref{eq:dNdh2}, with $g(\iota) \rightarrow 1$ since we are not accounting for inclination here. For $S \neq 0$, the Dirac delta does not vanish only for $f_0 \in B_k$, so the integral over frequencies is limited to that band. In the narrow-band approximation, we thus get
\barr
\frac{d N}{dS} &\approx& \Delta f ~ \int 4 \pi r^2 dr  d \mathcal{M}\frac{d n(r)}{d \mathcal{M} d f}\Big{|}_{f_k} \nonumber\\
&&~~~~~~  \times \delta \left(S - \frac{f_k}{\Delta f} \mathfrak{h}^2(\mathcal{M}, r, f_k) \right). \label{eq:dNdS-narrowband}
\earr
We now define the characteristic flux
\be
S_* \equiv \frac{f_k}{\Delta f } ~\frac{\mathfrak{h}^2(\mathcal{M}_*, r_*, f_k)}{(1 + z_*)^{4/3}}, \label{eq:S*}
\ee
and rewrite the Dirac delta function in Eq.~\eqref{eq:dNdS-narrowband} as 
\barr
&& \left(S - \frac{f_k}{\Delta f} \mathfrak{h}^2(\mathcal{M}, r, f_k) \right) \nonumber\\
 && = \delta\left( S - (1 + z)^{4/3} (r_*/r)^2 (\mathcal{M}/\mathcal{M}_*)^{10/3} S_*\right) \nonumber\\
 && = \frac3{10} \frac{\mathcal{M}}{S} \delta\left(\mathcal{M} - (r/r_*)^{3/5}(S/S_*)^{3/10} \frac{\mathcal{M}_*}{(1 + z)^{2/5}} \right).~
\earr
With this substitution, and now using our model for $d n/d \mathcal{M} d f$, we see that we can carry out the integral over $\mathcal{M}$, and arrive at
\barr
\frac{d N}{dS} &\approx& \frac{\pi^2}{16} \Delta f (\pi f_k)^{-11/3} ~\dot{n}_0 ~ r_*^3 ~\mathcal{M}_*^{-5/3} \frac{\sqrt{S_*}}{S^{3/2}} \nonumber\\
&& \times \int \frac{r dr}{r_*^2} e^{- r/r_* ( 1+ \sqrt{S/S_*})}.\label{eq:dNdS-prelim}
\earr
Computing the integral explicitly, and using Eq.~\eqref{eq:Nc-model}, we arrive at
\be
\frac{d N}{dS} \approx  2 \pi N_k~  \frac{\sqrt{S_*}}{S^{3/2}} \frac1{(1 + \sqrt{S/S_*})^2}.\label{eq:dNdS-model}
\ee
This distribution has the universal $\sim 1/S^{5/2}$ scaling for $S \gg S_*$, and scales as $\sim 1/S^{3/2}$ for $S \ll S_*$. It is straightforward to show that 
\be
\overline{h_{c, k}^2} = \int dS ~ \frac{d N}{dS} ~ S = 4 \pi~ N_k~ S_*, \label{eq:hck2-Nk-model}
\ee
which can also be checked from combining Eqs.~\eqref{eq:hc2-model}, \eqref{eq:Nc-model} and \eqref{eq:S*}. Note that Eq.~\eqref{eq:dNdS-model} is approximate only because of the narrow-band assumption, and can be readily generalized to a broad band by substituting $S_* \rightarrow S_* (f/f_k)^{4/3}$ and integrating Eq.~\eqref{eq:dNdS-prelim} over the finite frequency band.

\subsection{PDF of the characteristic strain squared}

We now compute the exact PDF $P(y \equiv h_{c, k}^2 / \overline{h_{c, k}^2})$ resulting from the source count distribution \eqref{eq:dNdS-model}, as a function of $y$ and $N_k$, using Eqs.~\eqref{eq:P(rho)-2}-\eqref{eq:mathcal(J)}. 

From Eqs.~\eqref{eq:dNdS-model} and \eqref{eq:hck2-Nk-model}, the rescaled source flux distribution $d N/ds$ takes the form of Eq.~\eqref{eq:dNds-F}, with
\be
\mathcal{F}(x) =  \left(1 + \sqrt{\frac{1}{4 \pi x}} \right)^{-2}, 
\ee
and the function $\mathcal{J}(\tau)$ of Eq.~\eqref{eq:mathcal(J)} is then
\barr
\mathcal{J}(\tau) = \frac1{4 \sqrt{\pi}} \int _0^{\infty} \frac{dt}{t^{5/2}} \frac{e^{i t} - 1 - it }{\left(1 + \sqrt{\tau/4 \pi t}\right)^2}.
\earr
This function can be written analytically in terms of the error function, sine and cosine integrals. The explicit expression is not particularly enlightening but is useful to speed up numerical evaluations. 

\begin{figure*}[ht]
\includegraphics[width=\linewidth]{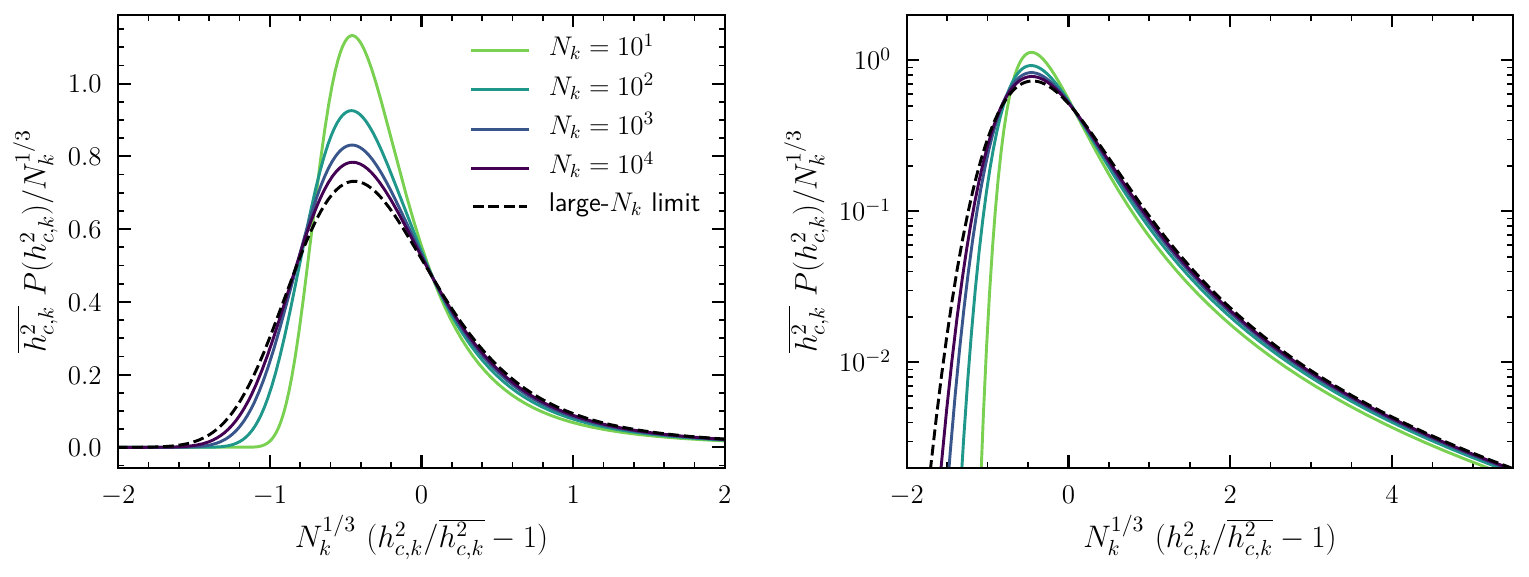}
\caption{Numerical PDF of the band-averaged characteristic strain squared for the simple SMBHB model described in Section \ref{sec:model}, as a function of $N_k$ (solid lines). The asymptotic large-$N_k$ limit presented in this paper is shown as a dashed line. For the ease of comparison, all PDFs are rescaled and shown as a function of the variable $x = N_k^{1/3}(h_{c, k}^2/\overline{h_{c, k}^2}-1)$.} \label{fig:comparison}
\end{figure*}

We obtain the PDF of $y = h_{c, k}^2/ \overline{h_{c, k}^2}$ for several values of $N_k \gg 1$ by explicitly computing the inverse Fourier transform of $e^{L(v)}$. We show the results in Fig.~\ref{fig:comparison}, where we compare the exact (numerical) PDFs against the large-$N_k$ limit given in Eqs.~\eqref{eq:main-1} and \eqref{eq:Airy-map}. For the ease of comparison, we plot all PDFs as a function of the rescaled variable $N_k^{1/3}(h_{c, k}^2/ \overline{h_{c, k}^2} -1)$, in which the large-$N_k$ limit takes a universal ($N_k$-independent) form. We see that the large-$N_k$ limit is indeed a good approximation for the exact PDFs, especially for $h_{c, k}^2 \gtrsim \overline{h_{c, k}^2}$. It very accurately reproduces the most probable value of $h_{c, k}^2$, as well as the characteristic widths of the numerical PDFs (although slightly overestimating them), for a broad range of $N_k$ values. In the low-$h_{c, k}^2$ regime, the exact PDFs do converge to the large-$N_k$ limit when $N_k$ is increased, but rather slowly, and with sharper cutoffs at $h_{c, k}^2 \lesssim \overline{h_{c, k}^2}$ than predicted by the analytic limit. This stems from the fact that the low-$h_{c, k}^2$ tail is more sensitive to the peak of the source flux distribution, in contrast with the high-$h_{c, k}^2$ tail, which is sensitive to its large-flux asymptote. 

\section{Practical application to Pulsar Timing Arrays (PTAs)} \label{sec:PTA}

In this section we discuss how to concretely apply the theorem we have presented above, and how it fits into -- and improves upon -- the existing framework for the analysis of PTA data.

\subsection{Analysis setup}

The fundamental data of a pulsar timing array (PTA) are the timing residuals time series of each pulsar, which we denote by $R_p(t)$. They can be written as a continuous Fourier series
\be
R_p(t) = \int_0^{\infty} df~ \left[C_p(f)\cos(f t) + S_p(f) \sin(f t) \right] . 
\ee
It is customary in PTA data analysis to approximate this continuous series by a discrete sum \cite{Lentati_13, Lentati_14}, 
\be
R_p(t) \approx \sum_k ~ \left[C_{p, k} \cos(f_k t) + S_{p, k} \sin( f_k t) \right].  
\ee
Note that this approximation is really only valid if $\Delta f \ll 1/T$, where $T$ is the total observing time, even though it is typically used with $\Delta f = 1/T$. We shall follow the usual PTA analysis convention here. We denote by $N_{\rm psr}$ the total number of pulsars.

\subsection{PDF of timing residuals from a Gaussian GWB}

In the limit of a truly infinite number of sources, the GWB strain is a Gaussian random field, with power spectrum proportional to $\overline{h_c^2}(f)$. 

Assuming timing noise (be it instrumental or intrinsic to the pulsar) is itself Gaussian, the coefficients $C_{p, k}, S_{p, k}$ are Gaussian distributed with zero mean. Different frequency bands are independent, and so are the sine and cosine amplitudes. For a given frequency band, the covariance of the cosine or sine amplitudes for pairs of pulsars $\{ p, q \}$ is then
\barr
&& \langle C_{p, k} C_{q, k} \rangle = \langle S_{p, k} S_{q, k} \rangle \nonumber \\
&&\approx \frac{\Delta f}{f_k} \left[\delta_{pq} \mathcal{N}(f_k) + (1 + \delta_{pq}) \mathcal{H}(\hat{p} \cdot \hat{q}) \frac{\overline{h_{c, k}^2}}{( 4 \pi f_k)^2} \right] \label{eq:covar},
\earr
where $\mathcal{N}(f)$ is the timing noise power spectrum per logarithmic frequency interval, $\overline{h_{c, k}^2}$ is the ensemble average of the band-averaged characteristic strain squared defined in Eq.~\eqref{eq:hk-def}, and $\mathcal{H}(\hat{p} \cdot \hat{q})$ is the well-known Hellings and Downs (hereafter HD) function\footnote{We assume $\mathcal{H}$ to be adequately normalized for Eq.~\eqref{eq:covar} to hold, so as to not have to unnecessarily keep track of factors of 2.} of the relative angle between pulsar directions \cite{Hellings_83}. Note that this covariance is only approximate, due to the discretization procedure. Note, also, that $h_c^2(f)/f^2$ would be a better quantity whose band-average to consider in the context of PTA data analysis, rather than $h_c^2(f)$ itself. It is straightforward to modify our derivation to account for the latter point: it suffices to redefine $h_{c, k}^2 \equiv f_k^3/\Delta f \int_{B_k} df h_c^2(f)/f^3$, and propagate the factors of $f^2$ throughout, noting that they make no difference in the narrow-band limit. 

For a given frequency bin $k$, we denote by $\bs{R}_k \equiv \{ C_{p, k}, S_{p, k} \}$ the $2N_{\rm psr}$-dimensional array of sine and cosine Fourier amplitudes of timing residuals, and by $P_{\rm HD}(\bs{R}_k | \overline{h_{c, k}^2})$ the Gaussian PDF with covariance given by Eq.~\eqref{eq:covar}, recalling that $\langle C_{p, k} S_{q, k} \rangle = 0$.

\subsection{Approximate PDF of timing residuals from a population of SMBHBs}

If SMBHBs are circular, they each emit at a single frequency, and the contributions of different frequency bands to the timing residuals are thus independent, as in the Gaussian case. In that limit we may consider the timing residuals from each frequency bin $\bs{R}_k$ one at a time, and obtain the PDF of the time-series $\{ R_p(t) \}$ by adequately convolving these PDFs. The independence of different frequency bands breaks down if SMBHBs are significantly eccentric \cite{Raidal_24}. We will not quantify this effect in this work, and only focus on the PDF of the $2N_{\rm psr}$ Fourier coefficients $\bs{R}_k$ for one frequency band at a time.

In principle, the exact $P(\bs{R}_k)$ resulting from GWs emitted by a population of discrete SMBHBs can be written explicitly as $2N_{\rm psr}$-dimensional Fourier transform, given explicitly in Ref.~\cite{Xue_25} for circular binaries. While the PDF can be computed for a single pulsar, it is computationally prohibitive to compute the joint $2N_{\rm psr}$-PDF of $\bs{R}_k$ for a realistic PTA.

As a consequence, the following Gaussian-mixture approximation is typically made: it is assumed that, for every \emph{realization} of the band-averaged characteristic strain squared $h_{c, k}^2$, the PDF of $\bs{R}_k$ is a Gaussian with HD correlations (and the adequate in-band noise contribution), that is
\be
P(\bs{R}_k) \approx \int d h_{c, k}^2~ P(h_{c, k}^2) ~ P_{\rm HD}(\bs{R}_k | h_{c, k}^2).\label{eq:P(R)-approx}
\ee
This approximation was implicitly used in NANOGrav's 15-year Astrophysical interpretation of the GWB \cite{Nanograv_23}, as well as in Ref.~\cite{Sato-Polito_25}. To our knowledge, the first paper that explicitly noted that Eq.~\eqref{eq:P(R)-approx} is only an \emph{approximation} was Ref.~\cite{Xue_25}. In that work, the authors explicitly showed that Eq.~\eqref{eq:P(R)-approx} is not mathematically equal to the exact PDF of timing residuals sourced by SMBHBs. They moreover showed numerically, using a realistic SMBHB model, that Eq.~\eqref{eq:P(R)-approx} is an accurate representation of the PDF of the timing residual amplitudes of a \emph{single pulsar}. This led them to conjecture that Eq.~\eqref{eq:P(R)-approx} may indeed be an accurate representation of the multi-pulsar timing residual joint distribution.

We now make two comments on the approximation \eqref{eq:P(R)-approx}, first on its regime of validity, and second on the accuracy with which $P(h_{c, k}^2)$ is usually computed.

First, we surmise that a necessary condition for the approximation Eq.~\eqref{eq:P(R)-approx} to be accurate is that the effective number of in-band sources $N_k$ is large relative to the number of pulsars. Indeed, the $2 N_{\rm psr}$ timing residual Fourier amplitudes are effectively determined by $6 N_k$ independent latent variables, which are, for each source, its strain amplitude $\mathfrak{h}$, two angles determining its position on the sky, two angles specifying the direction of the angular momentum (inclination and polarization angles), and one angle determining its orbital phase at a reference time. If $6 N_k \gg 2 N_{\rm psr}$, the large latent space implies that the distribution of timing residuals is smooth and has broad support on their $2N_{\rm psr}$-dimensional space, and we may reasonably expect (though this still has to be demonstrated) that Eq.~\eqref{eq:P(R)-approx} is an accurate approximation. On the other hand, if $6 N_k \lesssim 2 N_{\rm psr}$, the distribution of timing residuals only has support on a $6N_k$-dimensional sub-manifold of the full $2N_{\rm psr}$-dimensional space. Such a localized distribution cannot be well approximated by a Gaussian nor a linear superposition thereof as in Eq.~\eqref{eq:P(R)-approx}. Put differently, the $2N_{\rm psr}$ observables would be strongly dependent, in a way that cannot be qualitatively captured by Eq.~\eqref{eq:P(R)-approx}. We note that the fact that Ref.~\cite{Xue_25} found Eq.~\eqref{eq:P(R)-approx} to be a good approximation for a \emph{single pulsar} is consistent with our conjectured requirement that $6 N_k \gg 2 N_{\rm psr}$, which is more readily satisfied with $N_{\rm psr} = 1$. 

\begin{figure*}
\includegraphics[width = \linewidth]{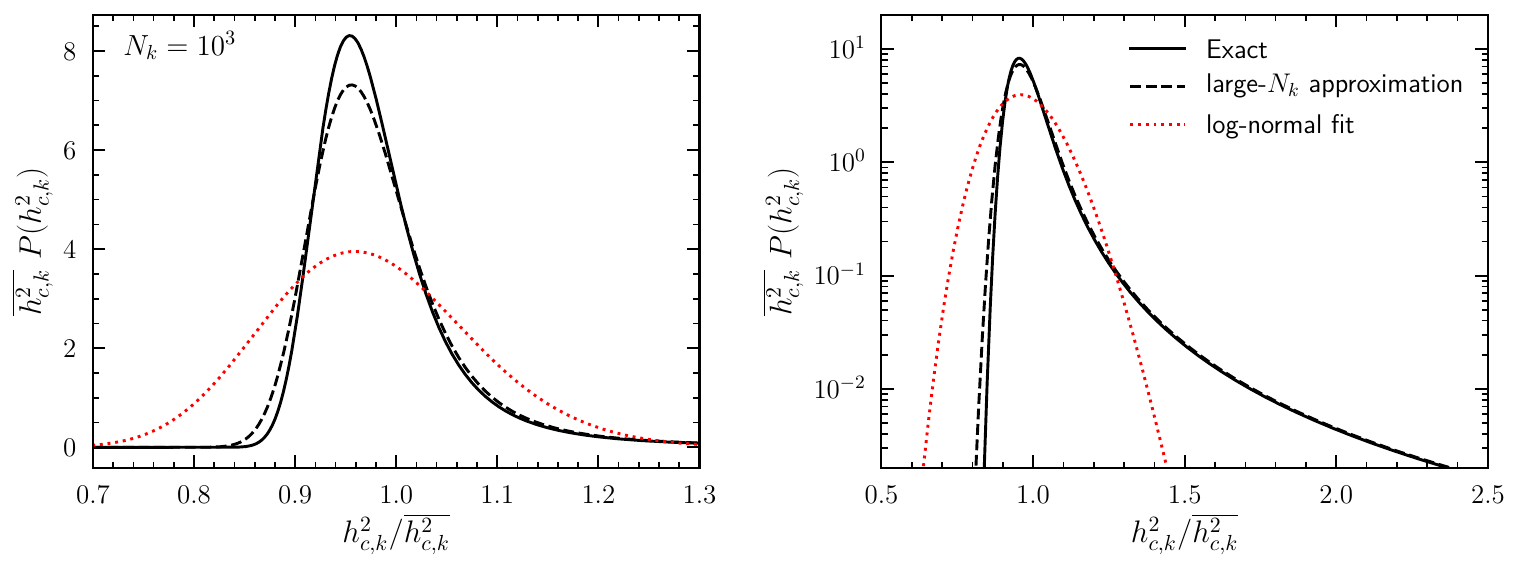}
\caption{Comparison of the exact PDF $P(h_{c, k}^2)$ computed numerically for the SMBHB model described in Section \ref{sec:model} (solid black), with the large-$N_k$ approximation presented in this work (dashed black), and a log-normal fit to the exact model (dotted red), as was done by the NANOGrav collaboration for the astrophysical interpretation of the GWB \cite{Nanograv_23}. The effective number of in-band sources was set to $N_k = 10^3$.}\label{fig:lognormal}
\end{figure*}

Second, let us note that, with the exception of simple semi-analytic models for the SMBHB population \cite{Sato-Polito_25, Xue_25}, the PDFs $P(h_{c, k}^2)$ are typically approximated numerically, by generating many realizations of a SMBHB population, and obtaining $h_{c, k}^2$ from Eq.~\eqref{eq:hk2-sum} with $g(\iota)$ set to unity (in other words, neglecting the inclination dependence of GW flux). In the original astrophysical interpretation of the NANOGrav 15-year data \cite{Nanograv_23}, Gaussian processes were trained on the median and rms of $\log_{10}(h_{c, k}^2)$ on a grid of binary parameters.  For each such parameter, the PDF $P(h_{c, k}^2)$ was then approximated as a log-normal distribution. This latter approximation makes the PDF used by NANOGrav especially inaccurate \cite{Laal_25}. We show this explicitly in Fig.~\ref{fig:lognormal}, where it can be seen that the log-normal fit to $P(h_{c, k}^2)$ is \emph{significantly} less accurate than our large-$N_k$ approximation across all values of $h_{c, k}^2$, and especially for the large-$h_{c, k}^2$ tail. 

Recently, Ref.~\cite{Laal_25} has implemented a normalizing flow (NF) emulator for $P(h_{c, k}^2)$ that is trained on the entirety of the GWB strain ensemble distribution rather than only the (log) median and rms, and therefore produces much more accurate $P(h_{c, k}^2)$. While we cannot directly compare our large-$N_k$ PDFs with their numerically-generated PDFs, we may compare their accuracy by computing the Hellinger distance with the exact PDF, defined as
\be
H \equiv \left[\frac12 \int dh_{c, k}^2 \left(\sqrt{P_{\rm appr}(h_{c, k}^2)} - \sqrt{P_{\rm exact}(h_{c, k}^2)}\right)^2\right]^{1/2}.
\ee
Using the simple model described in Section \ref{sec:model}, we find that the large-$N_k$ approximation and exact PDF have Hellinger distances $H = (0.21, 0.12, 0.06, 0.03)$ for $N_k = (10, 10^2, 10^3, 10^4)$, respectively. This should be compared to the 50\% percentile range $0.06 \leq H \leq 0.13$ reported by Ref.~\cite{Laal_25} for their NF emulator. We thus see that, for the specific model of Section \ref{sec:model}, the accuracy of our large-$N_k$ approximation is comparable to that of the NF emulator for $N_k \sim 10^2-10^3$, and superior to it for $N_k \gtrsim 10^3$. 

\subsection{Practical use of the theorem}

In summary, we have argued the following points in the previous section:\\
$\bullet$ The PDF \eqref{eq:P(R)-approx} that is routinely used for PTA data analysis is only an approximation, that we can reasonably expect to be accurate in the large-source-count regime, and that is almost certainly inaccurate for $N_k \lesssim N_{\rm psr}$.\\
$\bullet$ Current numerical techniques used to compute $P(h_{c, k}^2)$ for a given set of SMBHB population parameters are, at best, just as accurate as our large-$N_k$ approximation, especially for $N_k \gtrsim 10^2$ which is the physically relevant regime for the lowest frequency bins currently accessible to PTAs. 

We therefore propose to simply approximate $P(h_{c, k}^2)$ with the large-$N_k$ analytic solution $P(h_{c, k}^2|\overline{h_{c, k}^2}, N_k)$ provided in Eq.~\eqref{eq:main-1}-\eqref{eq:Airy-map}, which we argued should be just as accurate as existing implementations, while being extremely simple. In more detail, we propose to expand the current PTA analyses in the following ways:

$(i)$ The typical phenomenological analysis of PTA data consists in fitting a power-law for the mean characteristic strain squared, 
\be
\overline{h_{c, k}^2} = A_{\rm ref}^2 ~(f_k/f_{\rm ref})^{-\alpha},
\ee
where $\alpha = 4/3$ corresponds to circular binaries inspiralling through GW emission alone. The data is usually either analyzed while varying $A_{\rm ref}$ alone while fixing $\alpha = 4/3$, or varying both amplitude and slope. We propose that this analysis is expanded by accounting for a large but finite effective number of sources, also modeled as a power law of frequency,
\be
N_k = N_{\rm ref}~ (f_k/f_{\rm ref})^{-\beta},
\ee
using our large-$N_k$ approximation for $P(h_{c, k}^2)$. In the limit of circular binaries inspiralling through GW emission alone, $\beta = 11/3$ \cite{Sato-Polito_25}. One could then either let the two amplitudes and power-law indices float in a ``phenomenological"
analysis, or fix two of them at a time -- for instance, one could imagine a new ``fiducial" analysis where only $A_{\rm ref}$ and $N_{\rm ref}$ are allowed to vary, while $\alpha$ and $\beta$ are fixed to their fiducial values of $4/3$ and $11/3$ respectively.

$(ii)$ For the astrophysical interpretation of PTA data, we suggest to use SMBHB simulation pipelines (e.g.~\texttt{HOLDECK} \cite{Nanograv_23}) to extract the \emph{mean} characteristic strain squared $\overline{h_{c}^2}(f)$ and the cubic shot-noise-strain scale $\overline{h_0^3}(f)$ we defined in Eq.~\eqref{eq:h0^3-def} (or directly its band-dependent version $\overline{h_{0, k}^3}$ defined in Eq.~\eqref{eq:h03-eccentric} for eccentric binaries), as a function of frequency and model parameters, enforcing smoothness in both. Extracting these two summary statistics should be much more immune to sample variance than extracting the full PDF $P(h_{c, k}^2)$. A minimal modification of the current workflow would be to train Gaussian processes on $\overline{h_c^2}$ and $\overline{h_0^3}$ -- rather than on the median and rms of $\log_{10}(h_c^2)$ -- and then use the large-$N_k$ PDF $P(h_{c, k}^2)$ instead of a log-normal approximation. This would carry the same computational cost, with significant gains in accuracy.

\section{Conclusions} \label{sec:conclusion}

We have derived a simple and universal analytic approximation for the PDF of the band-averaged characteristic strain squared $h_{c, k}^2$ resulting from a population of SMBHBs in the large-source-count limit. This PDF is given by the map-Airy distribution, reflected, shifted, and rescaled. The relative width of the distribution is approximately $N_k^{-1/3}$, where $N_k$ is the effective number of in-band sources. The latter is determined from the mean characteristic strain squared $\overline{h_{c, k}^2}$ and the cubic shot-noise strain scale $\overline{h_{0, k}^3}$, a new summary statistic determined from the properties of the SMBHB population at $z \ll 1$. We argued that this large-$N_k$ approximation is at least as accurate as the current state-of-the-art approaches used to infer the PDF of $h_{c, k}^2$ from suites of realizations of the SMBHB population, and we outlined how it can easily be incorporated within the PTA data analysis framework in order to more efficiently -- and sometimes more accurately -- model discreteness effects in the GWB.

At a basic level, the large-$N_k$ approximation relies on the universal inverse-square-law scaling of the GW energy flux, and on the assumption that the SMBHB abundance converges to a non-zero value in the local Universe. Our results thus apply to an \emph{arbitrary} population of SMBHBs, be they circular or eccentric, and regardless of the dominant orbital hardening mechanism. An interesting expansion of our work would be to derive the large-$N_k$ joint PDF of $h_{c, k}^2$ across multiple frequency bands, as they are not independent if the SMBHB population is significantly eccentric \cite{Raidal_24}. We defer this calculation to future work.

In the context of PTA data analysis, the PDF of $h_{c, k}^2$ is the main ingredient of the Gaussian-mixture approximation for the joint distribution of pulsar timing residuals resulting from a discrete population of SMBHBs. While our results stand independently of the use of this approximation, it remains to be demonstrated that this Gaussian-mixture framework -- even when using the exact PDF of $h_{c, k}^2$ -- is an accurate approximation of the true PDF of timing residuals, especially for a large number of pulsars. This is a nontrivial task, which we leave to a future study.

Finally, while our main focus has been the nanohertz GWB from SMBHBs relevant to PTAs, our results should also carry over to other parts of the GWB probed by space- and ground-based detectors. It would be interesting to investigate whether similar large-source-count limits and associated self-similar statistics can improve the characterization of unresolved gravitational-wave backgrounds, or foregrounds, at higher frequencies.

\section*{Acknowledgements}

I thank Tristan Smith, Liang Dai and Gabi Sato-Polito for useful comments on this draft and enlightening conversations about the GWB and PTAs. I thank Hardi Veermae for kindly pointing out that the integral form of the large-$N$ limit was first derived in the appendix of Ref.~\cite{Ellis_23}.

\appendix

\section*{Appendix: Analytic form for the Universal large-$N_k$ PDF}

The realization that Eq.~\eqref{eq:P-integral} is equivalent to Eq.~\eqref{eq:Airy-map} was inspired by a statement made in Ref.~\cite{Addario-Berry_19}, that a 3/2-stable random
variable with Laplace transform $e^{4 \sqrt{\pi}s^{3/2}/3}$ is the map-Airy distribution introduced in Ref.~\citep{Banderier_01}. No proof was given of this statement -- and it turns out it is only correct up to a rescaling, i.e.~the correct Laplace transform should be $e^{s^{3/2}/3 \sqrt{2}}$. This appendix is thus dedicated to explicitly proving that Eq.~\eqref{eq:P-integral} is exactly equivalent to the reflected map-Airy function, Eq.~\eqref{eq:Airy-map}.

We recall that the Airy function can be defined by its Fourier transform \cite{Weisstein_Airy}
\be
\textrm{Ai}(x) = \int_{-\infty}^{\infty} \frac{dz}{2 \pi} e^{ix z + i z^3/3}.
\ee
This implies that the reflected map-Airy distribution given in Eq.~\eqref{eq:Airy-map}, can be rewritten as 
\barr
\mathcal{P}_{\textrm{Eq}\eqref{eq:Airy-map}}(x) &=& \int_{-\infty}^{\infty}  \frac{dz}{\pi} F_x(z), \\
F_x(z) &\equiv& - (x + i z) e^{i x^2 z + i z^3/3 + 2 x^3/3}.
\earr
The function $F_x(z)$ is analytic across the complex plane, and thus integrates to zero on any closed contour. At large $|z|$, $F_x(z)$ decays exponentially provided $z$ is in one of three quadrants: $0 < \textrm{arg}(z) < \pi/3$, $2 \pi/3 < \textrm{arg}(z) < \pi$, or $4 \pi/3 < \textrm{arg}(z) < 5 \pi/3$. We define the contour comprised of the real axis, the two 45-degree lines stemming from $ix$, and closed at infinity. Since the contour closes in the region where $F_x$ decays exponentially, the integral of $F_x$ along the real axis is equal to its integral along the V-shaped path with vertex at $ix$. We parametrize this path by $z = i x \pm \sqrt{t}~ e^{ \pm i \pi/4}$, with $t \in (0, +\infty)$, where the minus and plus signs correspond to the incoming (left) and outgoing (right) branches, respectively. We then have
\barr
- (x + i z) dz &=& \pm \frac{dt}{2},\\
i x^2 z + i z^3/3 + 2 x^3/3 &=& - \frac{t^3}{3 \sqrt{2}} \mp i \left(xt + \frac{t^{3/2}}{3 \sqrt{2}}\right).~
\earr
This implies that the contributions from both branches are the complex conjugates of one other, and the total integral is thus twice the real part of either branch, which gives precisely Eq.~\eqref{eq:P-integral}.

\bibliography{PrhoGWB.bib}

\end{document}